\documentclass[aps, prl, twocolumn, superscriptaddress]{revtex4-1}
\usepackage{graphicx}
\usepackage{dcolumn}
\usepackage{bm}
\usepackage{natbib}
\usepackage{upgreek}
\usepackage{amsmath}
\usepackage{amsfonts}
\usepackage{amssymb}
\usepackage{color}

\newcommand*{\Ho}{Ho$_{x}$La$_{3-x}$Ga$_5$SiO$_{14}$}

\begin{document}

%{\bf\Large
\title{Unusual magnetoelectric effect in paramagnetic rare-earth langasite}

\author{L. Weymann}
\author{L. Bergen}
\author{Th. Kain}
\author{Anna Pimenov}
\author{A. Shuvaev}
\author{E. Constable}
\author{D.~Szaller}
\author{A. Pimenov}
\affiliation{Institute of Solid State Physics, Vienna University of
Technology, 1040 Vienna, Austria}
\author{B. V. Mill}
\affiliation{Moscow State University, 119991 Moscow, Russia}
\author{A. M. Kuzmenko}
\author{V. Yu. Ivanov}
\affiliation{Prokhorov General Physics Institute of Russian Academy of Sciences, 119991 Moscow, Russia}
\author{N. V. Kostyuchenko}
\affiliation{Institute of Solid State Physics, Vienna University of
Technology, 1040 Vienna, Austria}
\affiliation{Moscow Institute of Physics and Technology, 141700 Dolgoprudny, Russia}
\author{A.~I.~Popov}
\affiliation{Prokhorov General Physics Institute of Russian Academy of Sciences, 119991 Moscow, Russia}
\affiliation{National Research University of Electronic Technology, 124498 Zelenograd,  Russia}
\author{A. K. Zvezdin}
\affiliation{Moscow Institute of Physics and Technology, 141700 Dolgoprudny, Russia}
\affiliation{Prokhorov General Physics Institute of Russian Academy of Sciences, 119991 Moscow, Russia}
\author{A. A. Mukhin}
\affiliation{Prokhorov General Physics Institute of Russian Academy of Sciences, 119991 Moscow, Russia}
\author{M. Mostovoy}
\affiliation{Theory of Condensed Matter, Zernike Institute for Advanced Materials, 9747 AG Groningen, The Netherlands}

\date{\today}

%\pacs{75.85.+t --- Magnetoelectric effects, magnetoelectric materials, rare-earth ions, crystal-field parameters, langasite}
%\small{\textbf{Keywords:} magnetoelectric materials, rare-earth ions, crystal-field parameters, aluminum borates}

\maketitle
%%%%%%%%%%%%%%%%%%%%%%%%%%%%%%%%%%%%%%%%%%%%%%%%%%%%%%%%%%%%%%%%%%%%%%%%%%%%%%%%%%%%%%%%%%%%%%%%%%%%%%%%%%%%%%%%%%%%%%%%%%%%%%%%%%%%%%%%%%%%%%%%%%%%%%%

\textbf{
Violation of time reversal and spatial inversion symmetries has profound consequences for elementary particles and cosmology.
Spontaneous breaking of these symmetries at phase transitions gives rise to unconventional physical phenomena in condensed matter systems, such as ferroelectricity induced by magnetic spirals, electromagnons, non-reciprocal propagation of light and spin waves, and the linear magnetoelectric (ME) effect -- the electric polarization proportional to the applied magnetic field and the magnetization induced by the electric field.
Here, we report the experimental study of the holmium-doped langasite, \Ho, showing a puzzling combination of linear and highly non-linear ME responses in the disordered paramagnetic state: its electric polarization  grows linearly with the magnetic field but oscillates many times upon rotation of the magnetic field vector.
We propose a simple phenomenological Hamiltonian describing this unusual behavior and derive it microscopically using the coupling of magnetic multipoles of the rare-earth ions to the electric field.}

%Violation of time reversal and spatial inversion symmetries has profound consequences for elementary particles and cosmology.
%
%Spontaneous breaking of these symmetries at phase transitions gives rise to unconventional physical phenomena in condensed matter systems, such as ferroelectricity induced by magnetic spirals, non-reciprocal propagation of light and spin waves, and the linear magnetoelectric (ME) effect -- the electric polarization proportional to the applied magnetic field and the magnetization proportional to the applied electric field.
%
%Here, we report the experimental study of the holmium-doped langasite, \Ho,  showing a puzzling combination of linear and highly non-linear ME responses in the disordered paramagnetic state:  its  electric polarization, $P$, oscillates many times upon the rotation of the applied magnetic field vector, $\mathbf{H}$,  but grows linearly with $|\mathbf{H}|$.
%
%We propose a phenomenological Hamiltonian  describing this unusual behavior and show how  it can be derived microscopically using the coupling of the electric field to magnetic quadrupole moments of the rare-earth ions, symmetry of this material and the hierarchy of energy scales in its excitation spectrum.

{\flushleft{\em Introduction}:}
Novel materials and physical effects are essential for the continuing progress in nanoelectronics ~\cite{manipatruni_natphys_2018}.
The coupling between electric and magnetic dipoles
in magnetoelectric and multiferroic materials enables electric control of magnetism that can significantly reduce power consumption of magnetic memory and data processing devices ~\cite{bibes_nm_2008, Kleemann_jap_2013, Matsukura_nn_2015, cheong_nmat_2007, spaldin_nmat_2019}.

The linear ME effect~\cite{landau_book8,dzyaloshinskii_jetp_1959,fiebig_jpd_2005} occurs in centrosymmetric crystals, such as Cr$_2$O$_3$~\cite{astrov_jetp_1960} and LiCoPO$_4$~\cite{rivera_fe_1994}, in which both time reversal and inversion symmetries are spontaneously  broken by an antiferromagnetic spin ordering inducing toroidal, monopole and other unconventional multipole moments~\cite{spaldin_jpcm_2008}. This effect allows for control of antiferromagnetic domains and gives rise to unidirectional light transmission at THz frequencies associated with electromagnons ~\cite{pimenov_nphys_2006, zimmermann_nc_2014, kocsis_prl_2018}.

However, the linear ME effect in centrosymmetric crystals requires a perfect match between magnetic and lattice structures, which narrows the pool of  ME materials substantially. Symmetry requirements are less stringent for magnets with chiral crystal lattices,  such as boracites~\cite{sannikov_jetp_1997}, copper metaborates~\cite{khanh_prb_2013} and the skyrmion material, Cu$_2$OSeO$_3$ ~\cite{seki_sci_2012}, in which diverse magnetic orders exhibit magnetoelectric behavior.

A bi-linear ME effect -- electric polarization proportional to the second power of the applied magnetic field -- does not even require time-reversal symmetry breaking and can occur in the paramagnetic state of chiral magnets through a variety of miscroscopic mechanisms\cite{ascher_phila_1968, schmid_spie_2000}. This higher-order effect is by no means weak: the magnetically-induced electric polarization in the recently studied holmium hexaborate is much higher than the polarization of linear magnetoelectrics at comparable magnetic fields~\cite{liang_prb_2011}.

The Ho-doped langasite studied in this paper does not seem to fit into this classification of ME effects. Langasite, La$_3$Ga$_5$SiO$_{14}$, and related compounds  have been synthesized as early as the 1980s~\cite{kaminskii_im_1982, kaminskii_izv_1983} and actively studied since then~\cite{mill_proc_2000} for their interesting non-linear optical, elastic and piezoelectric \cite{andreev_jetpl_1984, fritze_apl_2001} properties and applications in acousto- and electro-optics~\cite{stade_crt_2002, kong_jcg_2003}.
The non-centrosymmetric crystallographic space group P321 of langasites allows for optical rotation and piezoelectricity, but not for spontaneous electric polarization due to the presence of orthogonal three-fold and two-fold symmetry axes.
The study of magnetoelectric properties was focused on Fe-substituted langasites showing non-collinear periodically modulated orders of Fe spins~\cite{marty_prb_2010}.
The interplay between magnetic frustration, structural and spin chiralities gives rise to a rich variety of multiferroic behaviors ~\cite{loire_prl_2011, ramakrishnan_nqumat_2019}.

The Ho-doped langasite does not order down to lowest temperatures. Yet, its magnetically-induced electric polarization oscillates 6 times when the direction of the magnetic field rotates through 360$^\circ$ in the $ab$-plane, which is indicative of a highly non-linear magnetoelectric response proportional to (at least) the sixth power of the magnetic field. On the other hand, the magnitude of the induced polarization grows linearly with the  applied field, which seems to be at odds with the absence of any magnetic ordering in this material. This magnetoelectric behavior is very different from that of classical magnetoelectrics, such as Cr$_2$O$_3$. We show that it can be understood on the basis of langasite symmetry and a hierarchy of energy scales in the spectrum of the magnetic Ho ion.

\begin{figure*}
\begin{center}
\includegraphics[width=0.99\textwidth, clip]{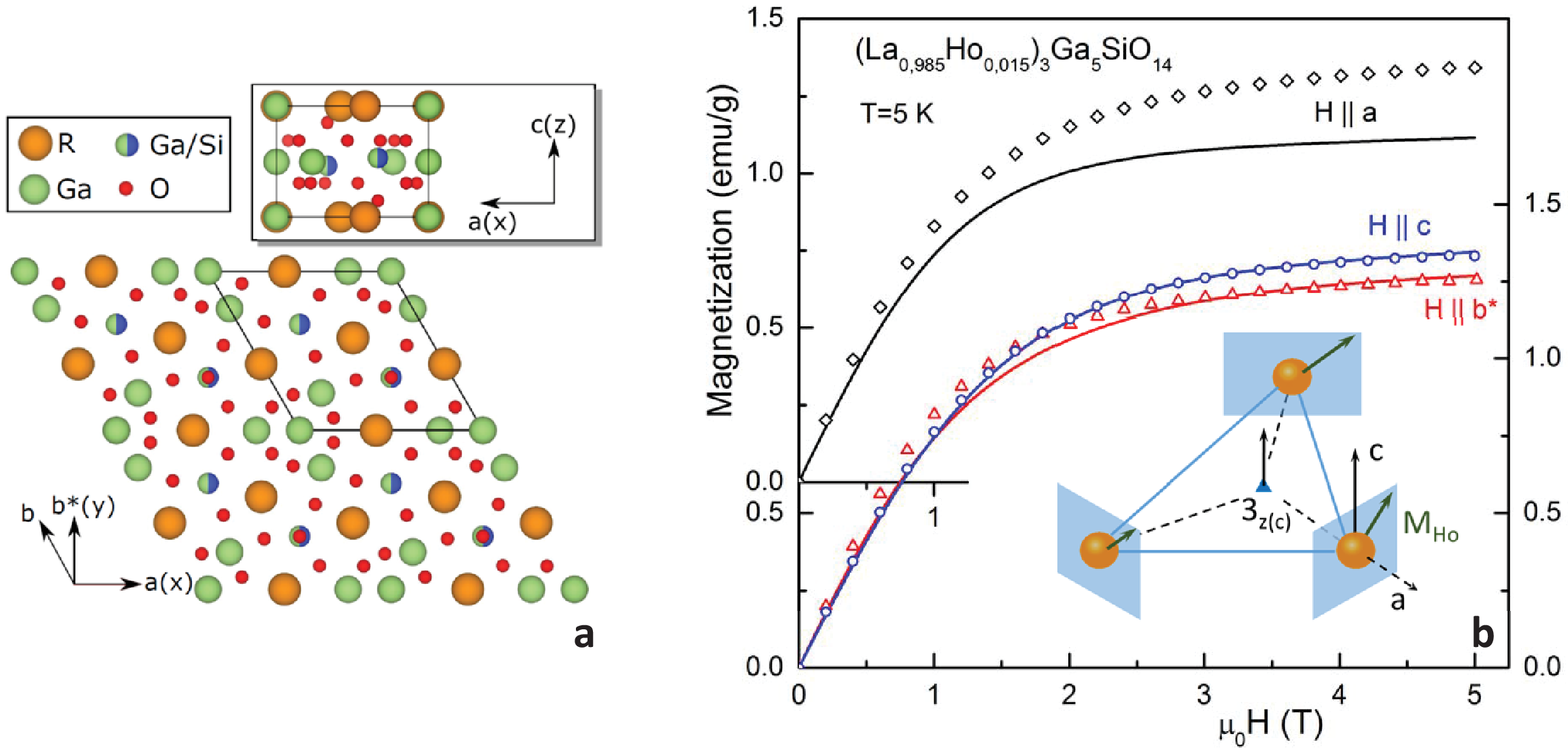}
\end{center}
\caption{\label{fig1} \textbf{Crystal structure and magnetism in rare-earth langasite.} \textbf{a}. The distorted Kagome lattice of the rare-earth ions in R$_3$Ga$_5$SiO$_{14}$. About 1.45\,\% of the rare-earth sites (orange) are occupied by the holmium ions. The unit cell is shown by thin solid lines. \textbf{b}. Experimental (symbols) and theoretical (solid lines) magnetization curves of Ho$_{x}$La$_{3-x}$Ga$_5$SiO$_{14}$ ($x=0.043$) in external magnetic fields parallel to the  $c$-axis at different temperatures as indicated. Solid lines - experiment, dashed lines - model accounting for a ground Ho$^{3+}$ quasidoublet and a Van-Vleck contribution of excited crystal-field states. The inset shows the local magnetic moments of Ho-ions in the saturation regime.}
\end{figure*}

{\flushleft{\em Results}:}
The samples used in this work are diluted rare-earth langasites, Ho$_{x}$La$_{3-x}$Ga$_5$SiO$_{14}$, with $x=0.043 \pm 0.005$.
The crystal structure of R$_3$Ga$_5$SiO$_{14}$ langasites~\cite{bordet_jpcm_2006} is shown in Fig.\,\ref{fig1}\textbf{a}.
The rare-earth sites, R,  have a rather low C$_2$ symmetry with the rotation axis along the $a$-axis of the crystal, allowing for electric dipole moments on R-sites, which in the case of magnetic Ho ions can depend on an applied magnetic field.

 The field-dependence of magnetization (Fig.\,\ref{fig1}\textbf{b}) shows saturation at fields $\sim 1$ T at low temperatures, which suggests that the magnetism of Ho$^{3+}$ ($^5{\rm I}_8$) ions in langasites is dominated by the two lowest-energy levels split by the crystal field from other states of the $J=8$ multiplet.
The level splitting in this so-called non-Kramers doublet, $\Delta$, is  estimated to be $3.1$\,K.
As discussed below, based on symmetry arguments and magnetization fits, the Ising-type magnetic moments of Ho-ions are oriented perpendicular to the local symmetry axis, i.e. in the $X_i Y_i$ plane, forming an angle $\gamma$ of $\sim 30^\circ$ with the crystallographic $c$-axis (Fig.\,\ref{fig1}\textbf{b}).

\begin{figure*}
\begin{center}
  \includegraphics[width=0.99\textwidth,clip]{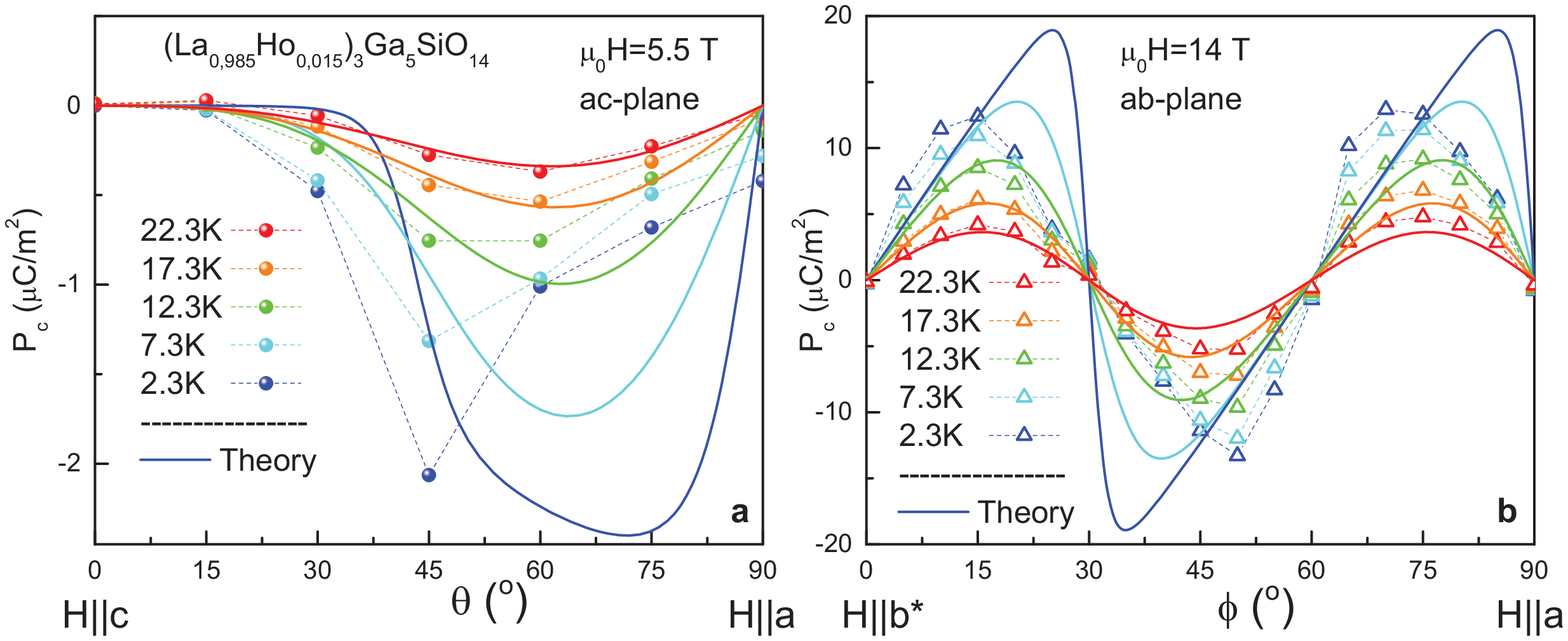}
\end{center}
\caption{\label{f_ang}
\textbf{Angular dependence of the static electric polarization along the $\mathbf{c}$-axis} \textbf{a}. $P_c$ vs the polar angle $\theta$ between the applied magnetic field, $\mathbf{H}$, and the $c$-axis, for  $\mathbf{H}$ rotated in the $ac$-plane at several temperatures. \textbf{b}. $P_c$ vs the azimuthal angle $\varphi$  for $\mathbf{H}$  rotated in the $ab$-plane. The modulation with the $60^\circ$ period is indicative of a high-order ME effect  (see Eq.\,(\ref{eq_p6})). Experimental data are shown with symbols, and dashed lines are calculated within the model described in the text.}
\end{figure*}

\begin{figure*}[tbp]
\begin{center}
\includegraphics[width=0.99\textwidth,clip]{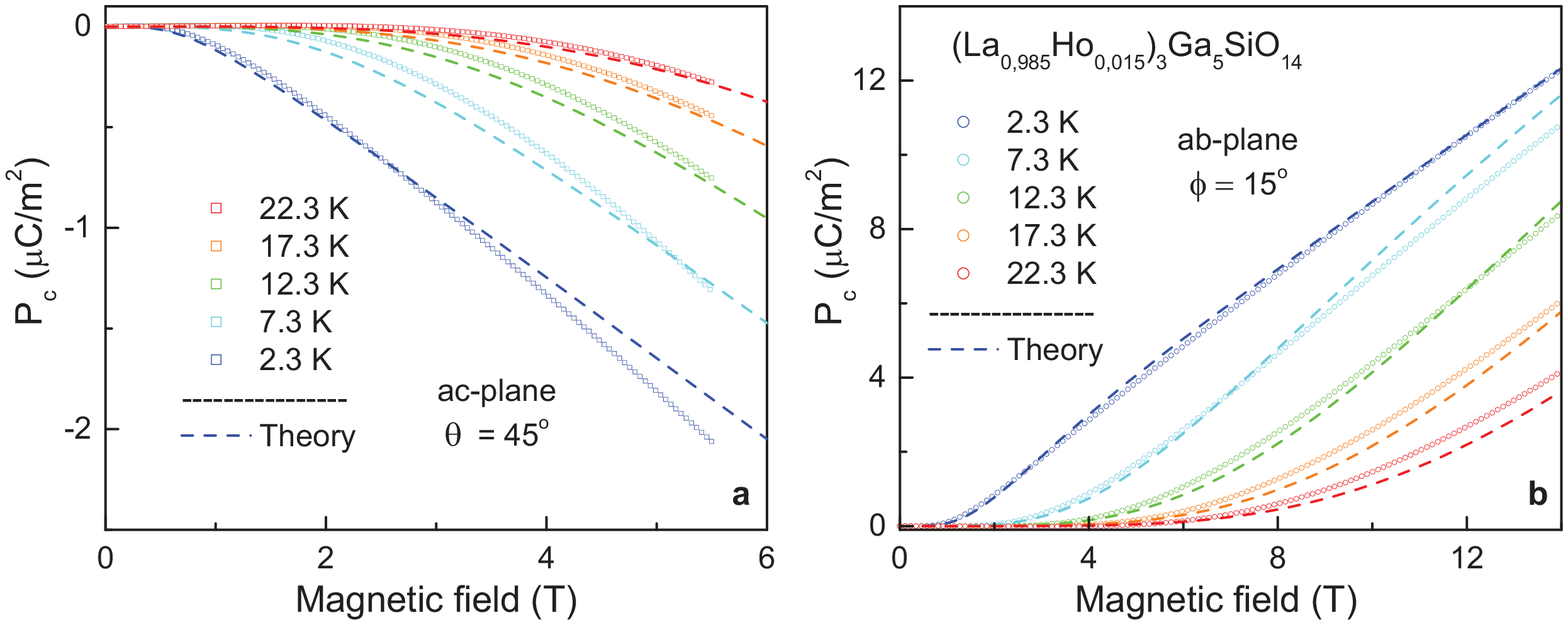}
\end{center}
\caption{\label{f_pol}
\textbf{Linear magnetoelectric effect in Ho-langasite.} \textbf{a}. Electric polarization for fields in the $ac$-plane with $\theta = 45^\circ$ (see also Fig.~\ref{f_ang}\textbf{a}). Solid lines are experimental data, dashed lines are results of theoretical calculations. \textbf{b}. Magnetic field dependence of the electric polarization, $P_c$, for applied magnetic fields rotated in the $ab$ plane through $15^\circ$ from the $b^*$-axis (see also Fig.~\ref{f_ang}\textbf{b}).
}
\end{figure*}

%\subsection*{Magnetoelectric effect}

 Figure~\ref{f_ang} shows the angular dependence of the electric polarization along the $c$-axis, $P_c$, for the magnetic field vector, $\mathbf{H}$, rotated in the $ac$- and $ab$-planes.
 The complex angular dependence is indicative of a high-order ME effect resulting from the C$_3$ symmetry of langasite.
 Although the expansion of the magnetically-induced electric dipole of the Ho-ion in powers of $H$ begins with terms $\propto H^2$, the bilinear ME effect is cancelled in the sum over the dipole moments of the Ho dopants in the three La sublattices.

The lowest-order phenomenological expression for $P_c$ allowed by C$_3$ and C$_2$ symmetries is
\begin{equation}
P_{z}^{(4)} = \alpha (T) H_xH_z (H_x^2-3H_y^2) ,
\label{eq_p4}
\end{equation}
where we use the Cartesian coordinates with $\hat{\mathbf{x}} =
\hat{\mathbf{a}}$, $\hat{\mathbf{y}} = \hat{\mathbf{b}}^\ast$,  $\hat{\mathbf{z}} = \hat
{\mathbf{c}}$ and $\alpha (T)$ is  a material constant.
 Fourth-order ME effect in crystals with trigonal crystal symmetry was discussed earlier~\cite{zvezdin_jetpl_2005, zvezdin_jetpl_2006, ivanov_jetpl_2017}.
Equation (\ref{eq_p4}) implies that $P_z \propto \sin^2 \theta \sin 2\theta \cos 3\varphi $, where $\theta$ and $\varphi$ are, respectively, the polar and azimuthal angles describing the magnetic field direction.
This angular dependence is in good agreement with the results of experimental measurements at high temperatures shown in  Fig.~\ref{f_ang}\textbf{a}, whereas at low temperatures the experimental $\theta$-dependence becomes more complex, indicating substantial higher-order contributions to the ME effect.

The fourth-order ME effect described by Eq.\,(\ref{eq_p4}) gives zero electric polarization for a magnetic field in the basal $ab$ plane.
However, the experimentally measured electric polarization shown in
Fig.~\ref{f_ang}\textbf{b} is of the same order as that for out-of-plane fields.
The electric polarization shows a periodic dependence on the azimuthal angle, $\varphi$,  with the period of $60^\circ$ apparently resulting from a sixth-order ME effect.
The lowest-order expression for the electric polarization induced by an in-plane magnetic field is, indeed, of sixth order in $H$:
\begin{equation}\label{eq_p6}
P_{z}^{(6)} = \beta(T) H_xH_y (H_x^2 - 3H_y^2)(3H_x^2 - H_y^2) \propto \sin 6\varphi.
\end{equation}
The saw-tooth shape of the angular dependence of the polarization at low temperatures (Fig.\,\ref{f_ang}\textbf{b})  indicates contributions of ME effects of yet higher orders.

Equations (\ref{eq_p4}) and  (\ref{eq_p6}) imply that the electric polarization is proportional to the 4$^{\rm th}$ or 6$^{\rm th}$ power of the applied magnetic field, depending on the orientation of ${\mathbf H}$. Instead, at low temperatures we obseve a nearly linear dependence of $P_z$ for any field direction (see Fig.\,\ref{f_pol}).  Importantly, phenomenological Eqs.~(\ref{eq_p4}) and  (\ref{eq_p6}) only hold for weak magnetic fields, $|\mu_0 H | \ll {\rm max}(\Delta, k_{\rm B} T)$, $\mu_0$ being the  saturated magnetic moment of a Ho$^{3+}$ ion (these equations are obtained in Methods by expansion in powers of $H$). At low temperatures, their validity is limited to fields $\lesssim 1$ T
(see Fig.\,\ref{fig1}\textbf{b}). The saturation of the Ho magnetic moment at higher magnetic fields gives rise to higher-order harmonics in the observed angular dependence of the electric polarization.

%Importantly, phenomenological Eqs.~(\ref{eq_p4}) and  (\ref{eq_p6}) only hold for weak magnetic fields, $|\mu_0 H | \ll {\rm max}(\Delta, k_{\rm B} T)$, $\mu_0$ being the  saturated magnetic moment of a Ho$^{3+}$ ion (these equations are obtained in Methods by expansion in powers of $H$). At low temperatures, their validity is limited to fields $\lesssim 1$ T
%(see Fig.\,\ref{fig1}\textbf{b}). The saturation of the Ho magnetic moment at higher magnetic fields gives rise to higher-order harmonics in the observed angular dependence of the electric polarization.

Counterintuitively, the onset of a more complex $\theta$- and $\varphi$-dependence correlates with the crossover to a \textit{linear}
dependence of $P_z$ on the magnitude of the magnetic field (Fig.\,\ref{f_pol}).
In what follows we discuss theory behind this unusual ME effect.

In the absence of inversion symmetry, the rare-earth ions interact with the electric field, $\mathbf{E}$, through an effective dipole moment operator, $\hat{\mathbf{d}}$, which in the subspace of the ground-state multiplet can be expressed in terms of the quadrupole and higher-order magnetic multipole moments of the ions ~\cite{popov_prb_2013}. The spectrum of the low-energy states depends then on both electric and magnetic fields, which gives rise to a ME response of the rare-earth ions. The description of the ME effect in  Ho-langasite can be simplified by projecting the Hamiltonian describing the magnetoelectric behavior of the ground-state multiplet on the subspace  of the two states, $| A \rangle$ and $| B \rangle$,  forming the non-Kramers doublet (see Methods).

Alternatively, the Hamiltonian describing the ME coupling of the non-Kramers doublet can be directly deduced from symmetry considerations.
Within the unit cell of Ho-langasite, there are three equivalent rare-earth positions with local C$_2$ symmetry axes along the  $\hat{Z}_i$ ($i = 1,2,3$) directions related by $120^\circ$-rotations around the crystallographic $c$ axis.
At site 1, $\hat{Z}_1 \| a$, $\hat{Y}_1 \| c$ (along the $\mathrm{C}_3$-symmetry axis) and $\hat{X}_1 = [\hat{Y}_1 \times \hat{Z}_1]$.
In the local coordinate frame, the ME Hamiltonian is,
\begin{equation}
\begin{split}
\hat{H}_{\rm me} = - i
&
\left(
   g_{XZ} E_X H_Z + g_{ZX} E_Z H_X
+ g_{YZ}E_Y H_Z \right. \\
&
\left.
+ g_{ZY} E_Z H_Y
\right) \times
\left(
|A \rangle \langle B | - | B \rangle \langle A |
\right),
\end{split}
\label{eq:Hme}
\end{equation}
(the index $i=1,2,3$ labeling the rare-earth ion is omitted for simplicity). Equation (\ref{eq:Hme}) is a general symmetry-allowed expression provided one of the states forming the doublet is even and another is odd under $C_2 = 2_{Z_i}$, in which case the magnetic moment of the doublet lies in the $X_iY_i$ plane and thus can have a nonzero component along the crystallographic $c$ axis. If both states would transform in the same way under $2_{Z_i}$, the magnetic moment would be parallel to the $Z_i$ axis and the magnetic moments of all Ho ions would lie in the $ab$ plane, in disagreement with experiment.

 The $Y$-component of the polarization ($P \| c$) measured in our experiment, results from the third term in Eq.\,(\ref{eq:Hme}). For $E_X = E_Z = 0$ and $E_Y = E_c$, the total effective Hamiltonian in the subspace spanned by the $| A \rangle$ and $| B \rangle$  states is:
\begin{equation}\label{eq:Heff}
\hat{H}_{\rm eff} = - d_Y E_c \sigma_x - ((\bm{\mu} \cdot \mathbf{H}) + g_{YZ} E_c H_Z) \sigma_y + \frac{\Delta}{2} \sigma_z,
\end{equation}
where $(\sigma_{x},\sigma_{y},\sigma_{z})$ are Pauli matrices, $\Delta$ is the crystal field splitting, $\bm \mu = (\mu_X,\mu_Y,0)$ and $d_Y$ are, respectively, the (transitional) magnetic and electric dipole moments of the Ho$^{3+}$ ion. The form of $\hat{H}_{\rm eff}$ is constrained by $2_Z$ and time-reversal symmetries of the Ho ions. In addition, $3_c$ symmetry implies that the (real) coefficients, $d_Y$, $\mu_X$, $\mu_Y$ and $g_{YZ}$ are the same for all Ho-sublattices.

The free energy of the doublet, $f_i$ ($i = 1,2,3$), can now be easily calculated and the magnetically-induced electric polarization is given by
\begin{equation}\label{eq:Pc}
\langle P_c \rangle = - \frac{n_{\rm Ho}}{3}\sum_{i=1}^3 \left( \frac{\partial f_i}{\partial E_c} \right)_{E_c = 0} = g_{YZ} \frac{n_{\rm Ho}}{3}\sum_{i=1}^3 H_{Z_i} M_i,
\end{equation}
where $n_{\rm Ho}$ is the density of Ho ions and $M_i$ is related to the average magnetic moment of the $i$-th Ho ion by $\langle  \bm \mu_i \rangle = (\mu_X,\mu_Y,0) M_i$ and is given by
\begin{equation}
M_i = \frac{\tanh \beta \epsilon_i }{\epsilon_i} (\bm{\mu} \cdot \mathbf{H}_i),
\end{equation}
where  $+\epsilon_i$ and $-\epsilon_i$ are energies of the two states forming the doublet, $(\bm{\mu} \cdot \mathbf{H}_i) = \mu_X H_{X_i} + \mu_Y H_{Y_i}$ and $\beta = 1/{k_{\rm B} T}$. In zero applied electric field, $\epsilon_i = \sqrt{(\bm{\mu} \cdot \mathbf{H}_i)^2 + (\Delta/2)^2}$
(for details see the Methods section and Refs.\,[\onlinecite{zvezdin_jetp_1981, babushkin_jetp_1983, vedernikov_jetp_1987,vedernikov_jetp_1988}]). The resulting magnetic field-dependence of the electric polarization (dashed lines in Figs.~\ref{f_ang} and \ref{f_pol}) is in good agreement with experimental data.

{\flushleft {\em Discussion}:} Figure~\ref{f_pol} shows that the electric polarization is a linear function of the applied field strength in the broad range of temperatures and external magnetic fields. The region of linear effect agrees well with the regime of saturated magnetic moments shown in Fig.\,\ref{fig1}\textbf{b}. This can be understood by noting that Eq.\,(\ref{eq:Pc}) can be obtained from the ME coupling, $-g_{YZ} E_{c} H_{Z_i} M_i$, on the $i$-th Ho sublattice. At low temperatures, $\beta \epsilon_i \gg 1$, $M_i$ grows with the applied magnetic field and approaches unity at rather weak fields, $|\mu_0 H| \sim \Delta$, above which the magnetic moment saturates and the ME coupling becomes linear.

Importantly, the ME coupling Eq.\,(\ref{eq:Hme}) originates from the admixture of higher-energy excited states from the $J = 8$ multiplet to $| A \rangle$ and $| B \rangle$ in the presence of electric and magnetic fields (see Methods). The linear ME effect is a consequence of a relatively large energy separation, $W$, of the non-Kramers doublet from the higher-energy states: it is observed for $\Delta \ll |\mu_0 H| \ll W$, when the dipole magnetic moment of the doublet is saturated while the admixture of higher-energy states remains small.

There is an additional ME coupling that results from the matrix elements of the electric and magnetic dipole operators between the $| A \rangle$ and $| B \rangle$ states in Eq.\,(\ref{eq:Heff}) and does not involve high-energy excited states. This coupling, however, is proportional to second and higher powers of the electric field and does not contribute to the magnetically-induced electric polarization measured in zero applied electric field, which can be traced back to the fact that the electric dipole is even and magnetic dipole is odd under time reversal.

Next we explain how the linear dependence of $P_c$ on the magnetic field strength  can be  compatible with the complex dependence on the direction of the field (see Methods for more details).
The six-fold periodicity of the polarization with the azimuthal angle
is related to the presence of three rare-earth sublattices.
Each sublattice separately gives rise to oscillations with the period of $180^\circ$ (see Eq.\,(\ref{eq_pc1})), since the Ising-like Ho magnetic moment changes sign as the magnetic field rotates in the $ab$-plane.
Similar two-fold periodicity is observed in the magnetoelectric Co$_4$Nb$_2$O$_9$ with Heisenberg spins, where the N\'eel vector describing an  antiferromagnetic ordering of Co spins rotates together with the magnetic field vector~\cite{khan_prb_2017}.
In addition, the ME effect in Ho-langasite is a sum of the responses of the three kinds of Ho positions with magnetic moments rotated through 120$^\circ$ around the $c$ axis, which gives rise to oscillations with the period of $60^\circ$.

{\flushleft {\em Conclusions}:}
 We have shown that the magnetic and magnetoelectric properties of Ho-langasite may be understood taking into account the two lowest crystal-field levels of Ho$^{3+}$ and the interplay of the local and global symmetries.
Our theory reconciles a complex oscillating dependence of the polarization on the magnetic field direction with a linear dependence on its magnitude.
The electric polarization is enabled by the absence of inversion symmetry at the local Ho positions.
In weak magnetic fields, when the Zeeman energy is small compared to the energy splitting in the non-Kramers doublet, the polarization is described by Eqs.\,(\ref{eq_p4},\ref{eq_p6}) and is highly non-linear in $H$.
It becomes linear in the strong-field limit, when the induced magnetic moment of Ho ions saturates, but it remains a complicated function of the magnetic field direction.

Importantly, the doping of non-centrosymmetric materials by magnetic rare-earth ions at low-symmetry positions provides a route to a linear ME response that does not require special magnetic orders and crystal lattices, as it occurs already in the paramagnetic state.  Oscillatory dependence of  magnetic and magnetoelectric responses on the direction of the applied magnetic field is governed by crystal symmetry and is a generic property of such systems.

\subsection*{Methods}

\subsubsection*{Experimental}
Single crystals of doped holmium langasite \Ho~ were grown by the Czochralski technique. For electric polarization a plane-parallel plate with thickness $d = 2.13$\,mm  and area  $A = 106$\,mm$^2$ has been cut with the flat surface perpendicular to the $c$-axis (c-cut).

For the electric polarization experiments silver paste electrodes were applied to two surfaces of the sample parallel to the $ab$-plane. The polarization was measured using an electrometer adapted to a Physical Property Measuring System, with magnetic fields of up to 14\,T and temperatures down to 2\,K. By changing the orientation of the sample in the cryostat, the direction of the magnetic field relative to the crystal axes was adjusted.
The magnetization was measured using a commercial Vibrating Sample Magnetometer with magnetic fields up to 6\,T. In these experiments, smaller samples from the same batch were used.

\subsubsection*{Microscopic theory of ME effect in Ho-langasite}
%\iffalse
%The symmetry of the rare-earth environment in langasites is described by the %point group C$_2$. This group is a cyclic one and has two one-dimensional %representations with respect to the rotation angle $\pi$ around the two-fold %symmetry axis (local $z$-axis): symmetric $A$ and antisymmetric $B$. %Ho$^{3+}$  with an even number of f-electrons is a non-Kramers ion. The %crystal field with C$_2$ symmetry completely removes the degeneracy of the %ground $^5I_8$ multiplet that splits into  $2J+1=17$ singlets. Their wave %functions are transformed either via A or B irreducible representations of %the C$_2$ group and they can be chosen in such a way that the matrix %elements of the Zeeman Hamiltonian $V_H = \mu_Bg_L\mathbf{JH} $, become %completely imaginary ($\mu_B$ and $g_L =5/4$ are the Bohr magneton and the %Lander factor, respectively). Simultaneously, the matrix elements of the %magnetoelectric Hamiltonian, Eq.\,(\ref{eq_vme}) become real since the %angular momentum operator $\mathbf{J}$ is odd and quadrupole (multipole) %momentum operators are even under the time reversal.
%\fi

The symmetry of the  rare-earth environment in langasites is described by the point group C$_2$.
This cyclic group has two one-dimensional representations with respect to the rotation through an angle $\pi$ around the two-fold symmetry axis (local $Z_i$-axis): the symmetric, $A$, and antisymmetric, $B$, representations.
Ho$^{3+}$  with an even number of $f$-electrons is a non-Kramers ion.
The crystal field with C$_2$ symmetry completely removes the degeneracy of the ground $^5I_8$ multiplet that splits into  $2J+1=17$ singlets with the wave functions transforming  according to either A or B irreducible representations of the C$_2$ group.

In the absence of inversion, the rare-earth ions interact with the electric field, $\mathbf{E}$, through an effective dipole moment operator, $\hat{\mathbf{d}}$, which in the space of the  Ho$^{3+}$  ground state multiplet can be expressed in terms of the quadrupole and higher-order magnetic multipole moments of the ions~\cite{popov_prb_2013}:
\begin{equation}\label{eq_d}
\left\{
\begin{array}{ccl}
\hat{d}_X &=&a_X \hat{Q}_{XZ} + b_X \hat{Q}_{YZ},\\
\hat{d}_Y&=&a_Y \hat{Q}_{XZ} + b_Y \hat{Q}_{YZ},\\
\hat{d}_Z&=&a_Z \hat{Q}_{ZZ} + b_Z (\hat{Q}_{XX}-\hat{Q}_{YY})+c_Z \hat{Q}_{XY},
\end{array}
\right.
\end{equation}
where $\hat{Q}_{\alpha \beta} = \tfrac{1}{2} [\hat{J}_{\alpha} \hat{J}_{\beta} +  \hat{J}_{\beta} \hat{J}_{\alpha} - \tfrac{2}{3}J(J+1)\delta_{\alpha \beta}]$  is the quadrupole moment operator, $\hat{\mathbf{J}}=(\hat{J}_X, \hat{J}_Y, \hat{J}_Z)$ being the total angular momentum operator of the Ho$^{3+}$ lowest-energy multiplet, $a,b$ and $c$ are phenomenological constants and the higher-order multipoles are omitted. Equation\,(\ref{eq_d}) is written in the local coordinate frame, ($X_i,Y_i,Z_i$), and the index $i=1,2,3$ labeling the rare-earth ion is omitted for simplicity. The phases of the wave functions forming the ground-state multiplet can be chosen in such a way that the matrix elements of the electric dipole (even under time-reversal) are real, whereas the matrix  elements of the magnetic dipole (odd under time-reversal) are all imaginary.

The ground state of the Ho$^{3+}$ ion in the crystal field potential, $V_{\rm cf}$, is a quasi-doublet (two close-by singlets with the gap $\Delta \sim $\,2\,$ \mathrm{cm}^{-1}$) which are separated by the energy $W \sim 30-50$\,$ \mathrm{cm}^{-1}$ from the excited crystal-field states.
The wave functions, $|A \rangle$ and $| B \rangle$, of the quasi-doublet can belong either to the same or to different representations.
As discussed in the main text, the $c$-component of the magnetization measured in the experiment is nonzero only in the latter case.

The response of Ho$^{3+}$ ions to the applied electric and magnetic fields is described by
$\hat{V} = - \hat{\mathbf{d}} \cdot \mathbf{E} + \mu_{\rm B} g_{\rm L}\hat{\mathbf{J}}\cdot{\mathbf{H}}$,
where the first term describes the coupling of the effective dipole operator Eq.\,(\ref{eq_d}) to the electric field and the second term is the Zeeman coupling,  $\mu_{\rm B}$ and $g_{\rm L} =5/4$ being, respectively, the Bohr magneton and the Lande factor.
To describe the magnetoelectric response of the Ho-langasite, we project the total Hamiltonian,  $\hat{H} = \hat{V}+\hat{V}_{\rm cf}$, on the quasi-doublet subspace:
\begin{eqnarray}\label{eq6}
\nonumber  \langle i | \hat{H}|j\rangle= &\tfrac{1}{2}& (-1)^i \Delta \delta_{ij} + \langle i|\hat{V}|j\rangle -  \\ &-& \sum_{k\neq1,2} \langle i|\hat{V}|k\rangle\langle k|\hat{V}|j\rangle/W_k  \ ,	
\end{eqnarray}
 where $i,j = 1,2$ label the states of the non-Kramers doublet, $|A \rangle$ and $| B \rangle$ and the effect of the higher-energy states from the $^5I_8$ multiplet ($k = 3,4,\ldots,17$) is treated in the second order of  perturbation theory assuming $|\langle k|\hat{V}|i\rangle| \ll W$, and $W_k = E_k - \frac{(E_1+E_2)}{2}$.
We note that this approach can also be used to analyze the rare-earth contribution to magneto-elastic and optical properties.

The matrix elements of $\hat{V}$ in the quasi-doublet subspace are constrained by $C_2$  symmetry (cf. Eq.\,(\ref{eq:Heff})):
\begin{equation}\label{eq:V}
\left(
\begin{array}{cc}
V_{11} & V_{12}\\
V_{21} & V_{22}
\end{array}
\right) = - d_0 E_{Z_i} I - d_Z E_{Z_i} \sigma_z -  \mathbf{d}\!\cdot\! \mathbf{E}_i \sigma_x -  {\bm \mu}\! \cdot \!\mathbf {H}_i \sigma_y,
\end{equation}
where $I$ is the $2 \times 2$ identity matrix, $\mathbf{d}\!\cdot\! \mathbf{E}_i = \langle A|\hat{\mathbf{d}}\!\cdot\!  \mathbf{E}_i|B \rangle = d_X E_{X_i} +d_Y E_{Y_i}$ and
${\bm \mu}\!\cdot\! \mathbf{H}_i = -i \mu_{\rm B} g_J \langle A|\hat{\mathbf{J}}\!\cdot\!\mathbf{H}_i|B \rangle  = \mu_X H_{X_i} +\mu_Y H_{Y_i}$.
Due to the $3_c$ symmetry of the langasite crystal, the real coefficients $d_0$, $d_X$,$\mu_X$, etc are the same for all Ho sites, i.e. are independent of $i$. The phases of the wavefunctions $| A \rangle$ and $|B \rangle$ are chosen such that the matrix elements of the time-even electric dipole operator involve real matrices, $I$ and $\sigma_x$, whereas the time-odd magnetic dipole operator is proportional to $\sigma_y$ with imaginary matrix elements.

In the last term of Eq.\,(\ref{eq6}) we only leave the off-diagonal parts proportional to the product of the matrix elements of
$\hat{\mathbf{d}} \cdot \mathbf{E}$  and $\mu_{\rm B} g_{\rm L}\hat{\mathbf{J}}\cdot{\mathbf{H}}$ giving rise to the magnetoelectric coupling,
\begin{equation}
\begin{split}
\hat{H}_{\rm me} =
&
\left(
   g_{XZ} E_{X_i} H_{Z_i} + g_{ZX} E_{Z_i} H_{X_i}
+ g_{YZ}E_{Y_i} H_{Z_i} + \right. \\
&
\left.
+ g_{ZY} E_{Z_i} H_{Y_i}
\right) \sigma_y,
\end{split}
\label{eq:Hme2}
\end{equation}
(cf. Eq.\,(\ref{eq:Hme})). Here,  the coefficients $g_{XY}$, $g_{YX}$, etc. are real numbers. This interaction is invariant under  $C_2$ and time-reversal symmetries, and since it involves both the electric and magnetic dipole operators, the matrix elements of $\hat{H}_{\rm me}$ in the quasi-doublet subspace are imaginary.

Adding all contributions to the effective two-level  Hamiltonian and assuming $E_{X_i} =  E_{Z_i} = 0$ and $E_{Y_i} = E_c$ ($i=1,2,3$), we obtain Eq.\,(\ref{eq:Heff}). The energies of the states forming the non-Kramers doublet at the
$i$-th Ho$^{3+}$ site, obtained by diagonalization of the Hamiltonian Eq.\,(\ref{eq:Heff}), equal $\pm \epsilon_i$, where
\begin{equation}\label{eq:epsilon}
\epsilon_i = \sqrt{(d_Y E_c)^2 + (\bm{\mu} \cdot \mathbf{H}_i +g_{YZ} E_c H_{Z_i})^2 + (\Delta/2)^2}.
\end{equation}
The free energy is then given by
\begin{equation} \label{eq12}
 f =  - \tfrac{1}{3} n_{\rm Ho} k_{\rm B} T \sum_{i=1}^{3} \ln
 \left(2 \cosh \frac{\epsilon_i}{k_{\rm B} T}\right),
\end{equation}
where $n_{\rm Ho}$ is the density of Ho ions and $k_{\rm B}$ is the Boltzmann constant. The average magnetic moment measured at zero applied electric field is
\begin{equation}
\begin{split}
\mathbf{M}  = &
- \left(\partial f / \partial \mathbf{H}\right)_{E_c = 0} \\
 = &
 \tfrac{1}{3} n_{\rm Ho} \mu_0^2
\sum_{i=1}^3 \mathbf{m}_i(\mathbf{m}_i \cdot \mathbf{H})
 \frac{\tanh \left(\beta \epsilon_i \right)}{ \epsilon_i},
\end{split}
\end{equation}
where $\beta = \frac{1}{k_{\rm B} T}$, $\mu_0$ is the  magnetic moment of the non-Kramers doublet and $\mathbf{m}_i$ is a unit vector defined by
\begin{equation}
\mu_X \hat{X}_i + \mu_Y \hat{Y}_i =  \mu_0 \mathbf{m}_i, \quad i = 1,2,3.
\end{equation}

The relation between the unit vectors in the hexagonal, Cartesian and the three local frames (see Fig.\,\ref{fig1}a) is:
\begin{equation}
\left\{
\begin{array}{lcc}
\hat{Z}_1 = \hat{\mathbf{a}} & = & \hat{\mathbf{x}},\\
\hat{X}_1 = \hat{\mathbf{b}}^\ast & = & \hat{\mathbf{y}},\\
\hat{Y}_1 = \hat{\mathbf{c}} & = & \hat{\mathbf{z}}.
\end{array}
\right.
\end{equation}
On all three sites $\hat{Y}_i = (0,0,1)$ in the Cartesian coordinates, $\hat{Z}_{2,3} = (-\frac{1}{2}, \pm \frac{\sqrt{3}}{2},0)$ and $\hat{X}_{2,3} = ( \mp \frac{\sqrt{3}}{2},-\frac{1}{2},0)$.

The $c$-component of the electric polarization measured in our experiment is given by
\begin{equation} \label{eq14}
\begin{split}
 P_c  = & - \left( \partial f / \partial E_c\right)_{E_c = 0} \\
 = &
 \tfrac{1}{3} n_{\rm Ho} g_{YZ} \mu_0
\sum_{i=1}^3 H_{Z_i}(\mathbf{m}_i \cdot \mathbf{H})
 \frac{\tanh \left(\beta \epsilon_i \right)}{ \epsilon_i},
\end{split}
\end{equation}
Note that since the electric polarization is measured at zero applied electric field, $d_Y$ that enters into the expression for the energy levels Eq.\,(\ref{eq:epsilon}), drops out from the expression for $P_c$, in which $\epsilon_i = \sqrt{(\bm{\mu} \cdot \mathbf{H}_i)^2 + (\Delta/2)^2}$. The energy splitting in the non-Kramers doublet by both electric and magnetic fields (see Eqs.\,(\ref{eq:V}) and (\ref{eq:epsilon})) does not result in the magnetoelectric coupling linear in electric field, because the electric and magnetic dipoles transform with opposite signs under time reversal. The observed magnetoelectric effect originates solely from the admixture of higher-energy states  in applied electric and magnetic fields (the last term in  Eq.\,(\ref{eq6})). In addition, Ho ions carry a ``built-in'' electric dipole moment along the local $Z_i$ axis in the $ab$ plane, which  results from the first term in Eq.\,(\ref{eq:V}) and is independent of the magnetic field. These electric dipoles on three Ho sites are rotated   through $120^\circ$ around the $c$ axis with respect to each other and  give no net contribution to the electric polarization.

In the limit of strong fields and low temperatures, $|\bm{\mu} \cdot \mathbf{H}_i|  \gg \Delta, k_{\rm B} T$, when the magnetic moment saturates, the electric polarization,
\begin{equation}\label{eq_pc1}
P_c   = \tfrac{1}{3} n_{\rm Ho} g_{YZ} \sum_{i=1}^3  H_{Z_i}
\mbox{sign} (\bm{\mu} \cdot \mathbf{H}_i) ,
\end{equation}
shows a complex saw-tooth dependence on the direction of the magnetic field in the $ab$ plane (due to the sign-functions)
and, at the same time, a simple linear dependence on the strength of the magnetic field, in agreement with our experimental observations.

In the opposite limit of weak fields and high temperatures, $k_{\rm B} T \gg |\mu_0 H|,\Delta$,
\begin{equation}
P_c  \approx \frac{n_{\rm Ho} g_{YZ} \mu_0^3 H^4}{4 (k_{\rm B} T)^3} \sin \gamma \cos^2 \gamma  \cos \theta \sin^3 \theta \cos 3 \varphi,
\label{eq:PhighT}
\end{equation}
where $H$ is the magnitude of the magnetic field and  $\gamma = \arccos (\mu_Y/\mu_0)$, which agrees with Eq.\,(\ref{eq_p4}) obtained by symmetry  and gives $\alpha (T) \propto T^{-3}$. At zero temperature,  $(k_{\rm B} T)^3$ in Eq.\,(\ref{eq:PhighT})  is replaced by $\Delta^3/12$. For magnetic  field applied in the $ab$ plane, the electric polarization at high temperatures,
\begin{equation}
P_c  \approx \frac{n_{\rm Ho} g_{YZ} \mu_0^5 H^6}{240(k_{\rm B} T)^5} \cos^5 \gamma \sin ^6 \theta \sin \left(6 \varphi \right),
\end{equation}
is proportional to $H^6$ and oscillates 6 times as $\varphi$ varies by $2\pi$, in agreement with Eq.\,(\ref{eq_p6}) ($\beta (T) \propto T^{-5}$). At zero temperature,   $(k_{\rm B} T)^5 \rightarrow \Delta^5/90$.

\subsection*{Acknowledgments}

We thank Peter Kregsamer and Christina Streli for help in determining the actual Ho-content in our samples.
This work was supported by the Russian Science Foundation
(16-12-10531) and by
the Austrian Science Funds (W 1243, I 2816-N27, P 27098-N27).

\end{document}